\documentclass{ws-p8-50x6-00}

\newcommand{\ic}{\;\;\; ,}
\newcommand{\ip}{\;\;\;.}
\newcommand{\sfrac}[2]{\textstyle \frac{#1}{#2} \displaystyle}
\newcommand{\half}{\sfrac{1}{2}}
\newcommand{\mev}{\mbox{MeV}}

\def\newblock{}

\begin{document}

\title{Hadronization in the Chromodielectric Model}

\author{G.~Martens, C.~Traxler, U.~Mosel}

\address{Institut f\"ur Theoretische Physik, Universit\"at Giessen}

\author{T.~S.~Biro}

\address{Research Institute for Particle and Nuclear Physics, Budapest}  

\maketitle

\vspace{-0ex}


\section{Introduction}
\label{sec:introduction}
Quantumchromodynamics (QCD) is the widely accepted theory to
describe strong interactions between hadrons. This theory shows
the well-known behavior of \emph{asymptotic freedom}. Furthermore,
lattice calculation show a phase transition from the hadronic world
to a system of free moving quarks and gluons, the Quark-Gluon-Plasma
(QGP). Heavy--ion experiments at CERN-SPS, at recently started
BNL-RHIC and at the yet to come CERN-LHC with energies of
$\sqrt{s}=20, 200, 5500$ A GeV respectively are designed to study the
eventually formed QGP\@. But still there is lack of a dynamical
description of both the transitions from hadrons to quarks and
gluons and vice versa, derived from first principles from QCD\@.
In this talk I present a classical,
molecular-dynamical model, which contains explicitly the phenomenon of
confinement and a dynamical mechanism for the formation of hadrons out
of large system of quarks and
gluons.\cite{Traxler:1998bk,Loh:1996nh,Kalmbach:1993sp,Vetter:1995gp} 


\section{The chromodielectric model (CDM)}
\label{sec:chrom-model}
We start with the Lagrange--density originally invented by Friedberg
and Lee \cite{lee1_77,lee2_77} and intensively studied by several followers
\cite{wilets89,wilets88,wilets82,wilets87}.
\begin{eqnarray}
  \label{eq:friedberg_lagrange}
  {\cal L}\; &=&\;\bar{q} \, (i\gamma_\mu D^\mu - m)\, q\:
              -\:\sfrac{1}{4}\, \kappa (\sigma) \, 
                           F_{\mu\nu}^a F^{\mu\nu\, a}\nonumber{}\\
           &&\; + \:(\partial_\mu\sigma)(\partial^\mu\sigma)
             \: - \: U(\sigma) \ip
\end{eqnarray}
The first term describes quarks,
where $q$ is a Dirac spinor with color--, spinor-- and flavor--indices
being suppressed and $m$ is a mass--matrix for the different
quark--flavor. $D^\mu = \partial^\mu + igA^\mu$ is the covariant
derivative, describing the minimal coupling to the gauge fields
$A^\mu = \frac{\lambda^a}{2} A^{\mu\,a}$ with coupling constant $g$
and $\lambda^a,\,a = 1\ldots8$, being the Gell--Mann matrices.

The second term is the kinetic term for the gauge field in a medium,
mediated via a dielectric function $\kappa (\sigma)$. The color field
tensor is given by $F_{\mu\nu}^a = \partial_\mu A_\nu^a -
\partial_\nu A_\mu^a - g f^{abc} A_\mu^b A_\nu^c$, where the $f^{abc}$
are the structure constants of SU(3)$_c$ and one has $-\frac{1}{4}
F_{\mu\nu}^a F^{\mu\nu\, a} = \frac{1}{2} (\vec{E}^a\!\!\cdot\!\!
\vec{E}^a - \vec{B}^a\!\!\cdot\!\! \vec{B}^a)$. $\vec{E}$ and
$\vec{B}$ are the color--electric and --magnetic fields.

The last two terms introduce a scalar field $\sigma$ with a quartic
scalar potential $U(\sigma)$. This scalar field is designed to mimic
the long-range behavior of non-perturbative QCD and is therefore purely
classical. It acts like a medium as in classical electrodynamics but
with a dielectric constant $\kappa(\sigma) < 1$ \cite{lee88}. The
potential $U(\sigma)$ is adjusted to have a global minimum at the
vacuum expectation value (VEV) $\sigma = \sigma_v$ and a local minimum
at $\sigma = 0$. In the absence of color--fields, the scalar field
takes on its VEV everywhere. 

\vspace{-1ex}

\subsection{Confinement mechanism}
\label{sec:conf-mech}
The mechanism of confinement in the CDM is based on an interplay of the
color--fields and the $\sigma$--field via the dielectric function
$\kappa(\sigma)$ and the scalar potential which are shown
schematically in fig.~\ref{fig:U_and_kappa}. 
From the Lagrangian (\ref{eq:friedberg_lagrange}) one gets the
equations of motion for the color--field.
\begin{eqnarray}
  \label{eq:full_eom}
  [D_\mu, \kappa(\sigma) F^{\mu\nu}] &=& j^\nu \ic %
\end{eqnarray}
where $j^\nu = g\, \bar{q} \gamma^\nu \lambda^a q \lambda^a/2$
is the color--current of the quarks.\footnote{Note that this current
  is not conserved, due to the color--carrying gluons.}
In an Abelian approximation the equations for the color--fields reduce
to the usual Maxwell--equations $\partial_\mu (\kappa(\sigma)F^{\mu\nu\,a}) =
j^{\nu\,a}$. The crucial point of the model is the choice of 
$\kappa(\sigma)$, which is supposed to contain all non--Abelian effects.
It is unity in the absence of the
$\sigma$--field and it vanishes when the scalar field takes on its
VEV $\sigma_{v}$. If one
considers a color--charge--distribution $\rho$ with vanishing total
color projections (in the Abelian directions 3 and 8), which we call a
white cluster, a color--field is produced due  
to the Gauss--law $\vec{\nabla}\!\cdot\!(\kappa(\sigma) \vec{E}^a) =
\rho^a$. The field is only allowed where $\kappa(\sigma) > 0$ ($\sigma
< \sigma_v$).  
To suppress the scalar field costs an energy $U(0) = B$,  and the
vacuum exerts a pressure on the color--field. 
If the
transition from $\kappa = 1$ to $\kappa = 0$ is a rapid one, then one
is left with a well defined spatial region, where the scalar field
nearly vanishes and the color field is non--zero. All
color--field--lines start and end on charges inside this volume and
therefore there are no Van-der-Waals--like interactions to other white
clusters except for very short-ranged $\sigma$-effects.
For that reason, if eventually two white subclusters form
inside the cluster, they can separate from each other.

In addition, if the charge--distribution has a non-vanishing total
charge, the field--energy deposited in this configuration is
divergent, i.~e.~those configuration cannot be created. 

\subsection{Model equations}
\label{sec:equations}

As the confinement mechanism in our model depends solely on the
specific choice of the dielectric function $\kappa(\sigma)$ and the
scalar potential $U(\sigma)$, we can neglect the spin--dependences in
the quark--Lagrangian. Instead we replace it with a Lagrangian for
classical, spinless charged particles that are coupled to the
classical color--field.
\begin{eqnarray}
  \label{eq:spin_0_lagrange}
  {\cal L}_p &=& \sum_k m_k \sqrt{1-\dot{\vec{x}}_k}\: 
               \rho_N\!(\vec{x}-\vec{x}_k)
               - j^{\mu a} A_\mu^a \\ 
  \label{eq:current}
  j^{\mu a} &=& g \sum_k\, q_k^a \, u_k^\mu \,
               \rho_N\!(\vec{x}-\vec{x}_k)
\end{eqnarray}
with the color--charge $q_k^a$  and the 4--velocity $u_k^\mu$. In our
numerical realization we deal with extended particles with a fixed
Gaussian distribution
\begin{equation}
  \label{eq:extended_particle}
  \rho_N(\vec{x} - \vec{x}_k) = \left(2 \pi r_0^2\right)^{-3/2} 
                               e^{-(\vec{x}-\vec{x}_k)^2/(2r_0^2)} \ic
\end{equation}
where $\sqrt{<\vec{x}^2>} = \sqrt{3}r_0 = 0.7\,$fm is chosen to fit
the radius of the nucleon. 

The color--current (\ref{eq:current}) is  consistent with the
originally derived equations of motion (\ref{eq:full_eom}) only 
in an Abelian sub--group of SU(3)$_c$. This is related to
the fact, that the gluons in QCD carry color as well, whereas the
color--current carried by the gluons vanishes in the Abelian
approximation \cite{Loh:1996nh,elze89}.

To further simplify the numerical realization we neglect the magnetic
fields $\vec{B}^a$. This is exact for static problems and for string
like yoyo--excitations \cite{wilets95}. The two decoupled sets of
Maxwell--equations reduce basically to the Gauss--law for each
field. To summarize we now have the following equations of motion
for the particles, the (electric) color--field and the $\sigma$--field
\begin{eqnarray}
  \label{eq:model_equations}
  \dot{\vec{x}}_k &=& \frac{\vec{p}_k}{E_k}\\
  \dot{\vec{p}}_k &=& -q_k^a 
               \int d^3x \left(\vec{\nabla} \phi^a(\vec{x})\right)
               \rho_N(\vec{x} - \vec{x}_k)\\
  \vec{\nabla}\cdot
    \left(\kappa(\sigma) \vec{\nabla} \phi^a(\vec{x})\right) &=&
                -\rho^a(\vec{x})\\
    \frac{\partial^2\sigma}{\partial t^2} + U'(\sigma) &=& 
                \nabla^2\sigma + \half \kappa'(\sigma) 
                \left(\vec{\nabla} \phi^a(\vec{x})\right)\cdot
                \left(\vec{\nabla} \phi^a(\vec{x})\right) \ic
\end{eqnarray}
where the prime denotes differentiation with respect to $\sigma$ and
$\phi^a(\vec{x})$ is the electric potential which satisfies 
$\vec{E}^a = -\vec{\nabla} \phi^a(\vec{x})$, $a \in \{3,8\}$. 
We choose for the scalar potential 
$U(\sigma) = B + a \sigma^2 + b \sigma^3 + c \sigma^4$ with $B = (150
\mev)^4$, $a = (489.9 \mev)^2$, $b = -15901 \mev$,  $c = 163.1$ and
for the VEV $\sigma_v = 61.1\mev$. The dielectric function
is $\kappa(\sigma) = \left(\exp\left(\alpha(\frac{\sigma}{\sigma_v} -
    \beta)\right) + 1\right)^{-1}$, where $\alpha = 7$ and $\beta =
0.4$. The coupling constant is chosen to reproduce the
string--tension in an $q\bar{q}$--configuration and takes on the value
$\alpha_{\scriptscriptstyle S} = g/(4\pi) = 2$.

\vspace{-1ex}

\subsection{Classification of particles}
\label{sec:particles}

In QCD the quarks and gluons are represented as triplet-- and
octet--states  of SU(3)$_c$ respectively. In our classical simulation
we assign classical charges $q^a$ to the quarks. These charges are the
diagonal entries of the $\lambda^a, a\in{3,8}$. 
Due to our approximation we neglect the non--Abelian part of the
color--fields. Instead we treat these
6 gluon--fields as particles in the same formalism as the
quarks except that these particle--gluons carry both a color and an
anti-color. The corresponding charges for quarks, anti-quarks and gluons
are depicted in fig.~\ref{fig:colorscheme}. E.\/g.~the particle $r$ in that
scheme is a quark of color \emph{red} and color--charges 1 and
$1/\sqrt{3}$ with respect to the 3-- and the 8--field respectively.

As we have mentioned in section \ref{sec:conf-mech}, the dynamics of our
model forces the charged particles into white clusters and the
separation into white subclusters is allowed. We regard as hadrons
only white clusters which cannot be divided into smaller ones. It
turns out, that there is only a finite set of those \emph{irreducible
  white Clusters} (IWC). The IWCs consist either of a quark and an
anti-quark or of three quarks (anti-quarks), both with some gluon
admixture, or they consist only of gluons and can therefore easily be 
interpreted as mesons, baryons and glue-balls respectively.

For the particle masses we use constituent quark masses to fit the
low--lying hadronic spectrum. As our model does not depend on isospin,
we treat the $u$-- and the $d$--quark as degenerate particles with
the same mass. Only the light pion, which is assumed to be a
Goldstone--boson of chiral symmetry breaking, does not fit in our
constituent-mass--scheme and thus we do not incorporate pions in our
model. The quark masses are fixed to be $m_{u,d} = 400$MeV, $m_s =
550$MeV and $m_c = 1500$MeV. Thus the masses of the lowest hadronic states
are simply the sum of the quark masses of the IWC.
For the gluons we take a mass $m_g = 700$MeV to reproduce the lightest
glueball-mass of 1400MeV.

\vspace{-1ex}


\section{Hadronization}
\label{sec:hadronization}

\vspace{-0.3cm}

\begin{figure}[htbp]
    \begin{minipage}[t]{\textwidth}
      \begin{center}
        \epsfxsize=0.33\textwidth %
        \epsfbox{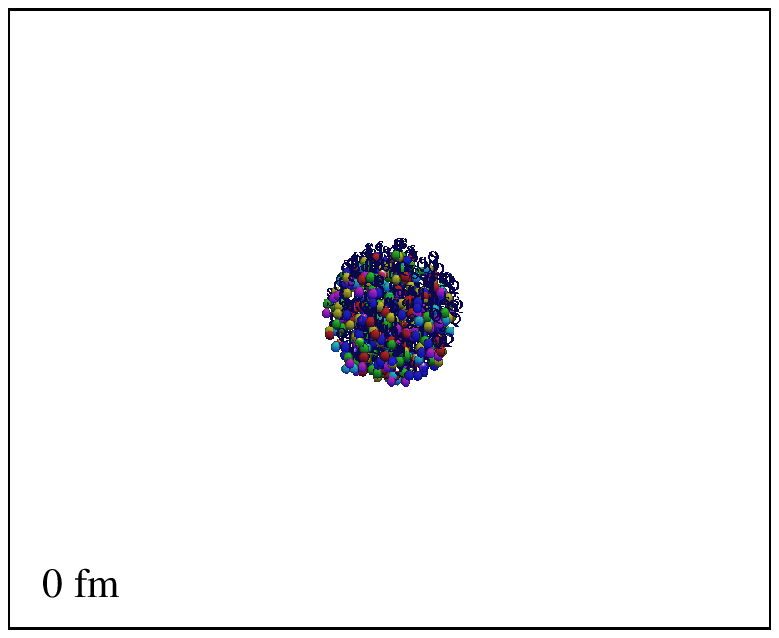} \hspace{-0.2cm}%
        \epsfxsize=0.33\textwidth %
        \epsfbox{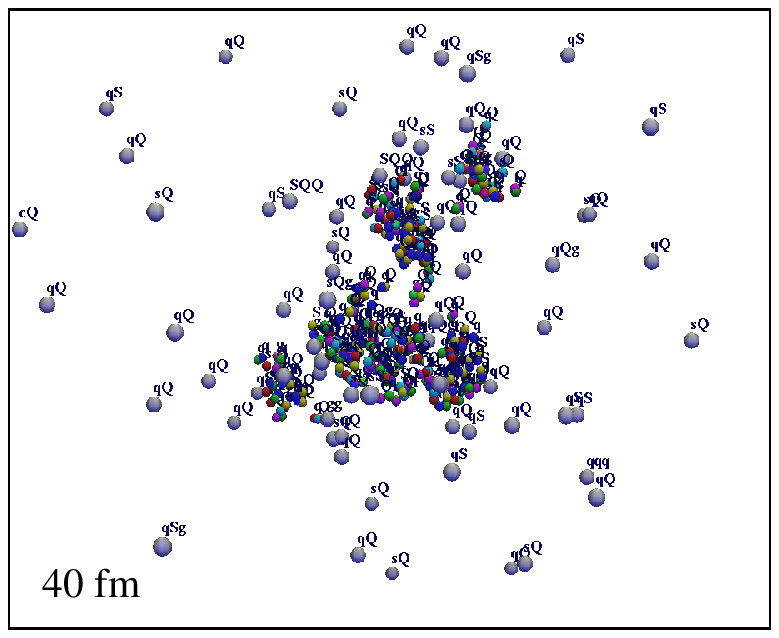} \hspace{-0.2cm}%
        \epsfxsize=0.33\textwidth %
        \epsfbox{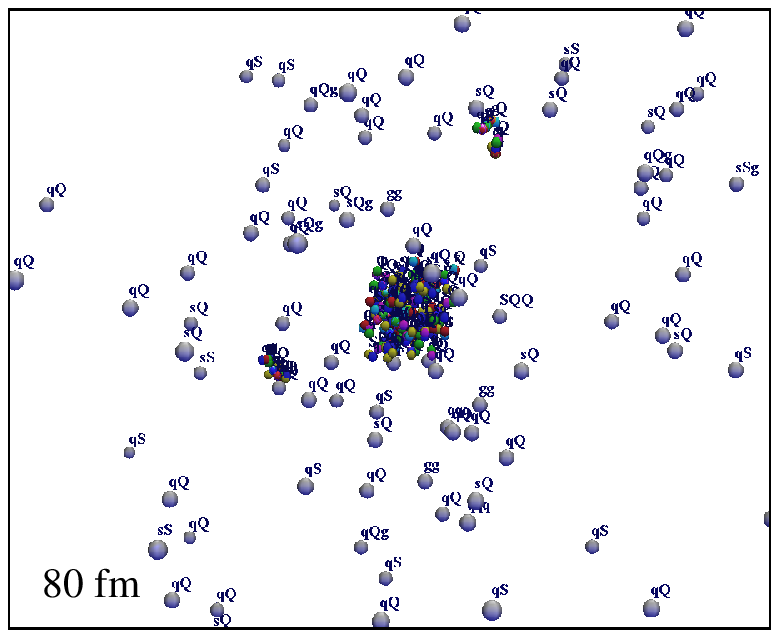} %
        \caption{Different timesteps of the hadronization
          process. The simulation starts with a  droplet of
          quarks and gluons and ends after about 200 fm left with
          IWCs only. The particle labels desribe the flavor content
          and grey shaded particles are already detected IWCs.
          \label{fig:hadron_scenario}}
      \end{center}
    \end{minipage}
\end{figure}

To simulate the hadronization out of a QGP, we start with an ensemble
of quarks and massive gluons, distributed homogeneously in
a sphere of radius $R = 4$fm in real space and according to a
Boltzmann--distribution with initial temperature $T_0 = 160$MeV in
momemtum--space. The relative number of different particles is given
through the distribution $N_i \propto d_i\,\exp\left(-m_i/T_0\right)
\left(\frac{m_i T_0}{2\pi}\right)^{3/2}$, where $d_i$ is a
(spin, isospin and color) degeneration factor. The colors are chosen
randomly with the constraint of overall color--neutrality. After
solving the Gauss--law for the color--fields in the first time step
the system is driven by the equations of motion
(\ref{eq:model_equations}). Due to the initial momenta the particles
tend to leave the system but are bound due to the formation of
color--strings. In this way the particles are reorganized to form in a
first step white clusters and finally only IWCs. This scenario is shown
in fig.~\ref{fig:hadron_scenario}. 
The hadrons have
invariant masses $M_{\mbox{iwc}}^2 = E^2 + \vec{P}^2$
with particle energies $E_i$ and momenta $\vec{p_i}$
\vspace{-0.5ex}
\begin{eqnarray}
  \label{eq:invariant_mass}
  E &=& \sum_i E_i + \int d^3x \left(
    \half \dot{\sigma}^2 + \half(\vec{\nabla}\sigma)^2 + U(\sigma)
    + \half \kappa(\sigma) \vec{E}^a\vec{E}^a
  \right)\\
  \vec{P} &=& \sum_i \vec{p}_i 
  - \int d^3x \dot{\sigma}\vec{\nabla}\sigma \ip
\end{eqnarray}

\begin{figure}[htbp]
  \begin{minipage}[t]{0.9\textwidth}
    \begin{center}
      \epsfysize=11pc %
      \epsfbox{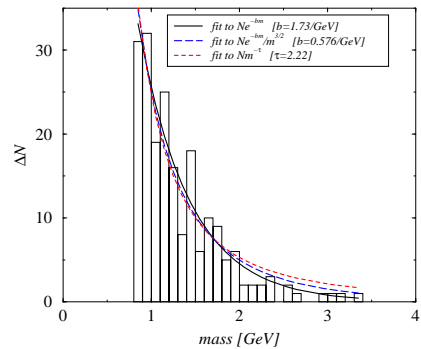} %
      \vspace{-0.5cm} 
      \caption{The mass distribution of the resulting
          hadrons. The 
          curves are fits to a Hagedorn distribution with $a=-3/2$ and
          $a=-3$ and $b = (1/T - 1/T_h)$ and to an inverse
          power--law--distribution.
        \label{fig:mass_distribution}}
    \end{center}
  \end{minipage} %
\end{figure}

\vspace{-1.5ex}
The resulting mass distribution is shown in
fig.~\ref{fig:mass_distribution}. The curves are fits to a Hagedorn
distribution $dn/dm \propto m^{a+3/2}\exp(-m(1/T-1/T_h))$ and to an
inverse power--law--distribution $dn/dm \propto m^{-\tau}$. If we
assume $T_h = 160$MeV and $a = -3$ or $a = -3/2$ in the Hagedorn case,
we get a hadronic temperature $T = 146$MeV and $T = 125$MeV
respectively. 


\begin{figure}[htbp]
  \begin{center}
    \begin{minipage}[t]{0.3\textwidth}
      \epsfysize=8pc %
      \epsfbox{Ukappa.eps} %
      \caption{The scalar potential $U(\sigma)$  and the
        dielectric function $\kappa(\sigma)$.} \label{fig:U_and_kappa}
    \end{minipage}\hfill
    \begin{minipage}[t]{0.65\textwidth}
      \begin{center}
        \epsfysize=8.5pc %
        \epsfbox{colors.eps} %
        \caption{The color scheme in the Abelian approximation. On the
          3(8)--axis is given the particle charge with respect to the
          3(8)--field. The particle on the two triangles (r, g, b) and
          ($\bar{r}$, $\bar{g}$, $\bar{b}$) represents the quarks and
          anti-quarks and those on the hexagon the 6 massive gluons.}
        \label{fig:colorscheme} 
      \end{center}
    \end{minipage}
  \end{center}
\end{figure}

\end{document}